# Observation of anomalous thermal Hall effect in altermagnets

Wenbo Wan [1,2*], Xu Zhang [1*✉], Yixuan Luo [3], Yanfeng Guo [3,4✉] and Shiyan Li [1,2,5,6✉]

[1] *State Key Laboratory of Surface Physics, and Department of Physics, Fudan University, Shanghai 200438, China*

[2] *Shanghai Research Center for Quantum Sciences, Shanghai 201315, China*

[3] *State Key Laboratory of Quantum Functional Materials, School of Physical Science and Technology, ShanghaiTech University, Shanghai 201210, China*

[4] *ShanghaiTech Laboratory for Topological Physics, ShanghaiTech University, Shanghai 201210, China*

[5] *Shanghai Branch, Hefei National Laboratory, Shanghai 201315, China*

[6] *Collaborative Innovation Center of Advanced Microstructures, Nanjing 210093, China*

Corresponding author. Email: shiyan_li@fudan.edu.cn (S.Y.L.); guoyf@shanghaitech.edu.cn (Y.F.G.); xuzhang_fd@fudan.edu.cn (X.Z.)

## Abstract

**Altermagnets, recently proposed as a third category of collinear magnets, combine the features of zero net magnetization in antiferromagnets and the spin splitting in ferromagnets. While abundant spectroscopic evidence for altermagnetism has been reported, experimental observation of the anomalous Hall effect, a hallmark of ferromagnetism, remains scarce. Here, we shift the paradigm from charge to heat carriers and report the systematic study of the thermal Hall effect in two representative altermagnet candidates, MnTe and CrSb. In both materials, we observe a pronounced anomalous phonon thermal Hall signal, with no electrical counterpart observed, attributed to the coupling of this distinctive magnetic structure with phonons. Our findings establish the anomalous phonon thermal Hall effect as an intrinsic feature of altermagnets, and provide a sensitive probe to identify this new kind of quantum magnets. The anomalous phonon thermal Hall**

**effect in altermagnets directly links the Néel vector to lattice vibrations, opening prospects for low-loss phononic devices and thermally readable memories.**

## Introduction

Collinear magnets are conventionally categorized into two main types: ferromagnets (FMs), where parallel spin alignment generates momentum-space band splitting and breaks time-reversal symmetry (TRS), and antiferromagnets (AFMs), in which antiparallel spin sublattices yield zero net magnetization. Recently, altermagnets (AMs) have been proposed as a third distinct form of collinear magnets, defined by alternating spin orientations in both real and reciprocal space, which combine characteristics of FMs and AFMs[1–4]. In altermagnets, long-neglected nonmagnetic lattice atoms, e.g., Te in MnTe and Sb in CrSb (see Fig. 1a and Fig. 3a), play a crucial role: the spin space group symmetries allow magnetic sublattices to be connected solely through rotation or mirror operations, enabling spin splitting even in the absence of spin–orbit coupling (SOC) and a finite net magnetization[1,2,5,6].

Since the introduction of altermagnetism, several materials have been identified as potential altermagnets, with $RuO_2$ recognized as the first candidate predicted to exhibit the anomalous Hall effect (AHE) that arises from crystal TRS breaking[7]. Although nonlinear Hall signals have been observed in $RuO_2/TiO_2$ (110) films[8], spin- and angle-resolved photoemission spectroscopy (ARPES) has not given any evidence for the expected spin splitting[9], leaving the validity of the proposed altermagnetic order in rutile $RuO_2$ under scrutiny. More recently, spin splitting has been observed in CrSb[10–13] and MnTe[14–18], yet clear detection of anomalous Hall responses remains elusive[19–25]. In CrSb, nonlinear Hall signals are attributed to multi-band effects[19–21], whereas in MnTe, AHE hysteresis loops are not only very weak but also exhibit strong sample dependence[22–25]. Note that for $RuO_2$, saturation of the AHE is observed without zero-field AHE because of limitation by symmetry[26,27]. Therefore, identifying genuine altermagnetic materials using multiple experimental techniques has become a key focus of current research.

Over past two decades, the thermal Hall effect (THE) in insulators has gained significant attention as a novel phenomenon[28–44]. According to these works, despite the absence of charge carriers, insulators subjected to a longitudinal temperature gradient can still generate a transverse temperature difference in the presence of magnetic field. Distinct origins of the THE have been proposed, including magnons, various exotic excitations, and phonons. Most strikingly, such THE has been observed in various trivial nonmagnetic insulators (including $SiO_2$, MgO and $MgAlO_2$) and semiconductors (including Si and Ge) with a scaling law of $|\kappa_{xy}| \sim \kappa_{xx}^2$, demonstrating a universal phonon THE (pTHE) in single crystals[45]. In this context, many previous exotic interpretations may not be needed, but the underlying physics of this universal pTHE is still under investigation[46,47].

In magnetic materials, spin degree of freedom can significantly influence the pTHE, causing deviation from the quadratic scaling behavior[45]. In collinear AFMs, the thermal Hall conductivity varies approximately linearly with magnetic field[34–36,39,44], similar to pTHE in nonmagnetic materials. By contrast, an anomalous THE has been observed in FMs[29,32,38], indicating spontaneous TRS breaking, while its origin—whether from magnons or phonons—remains unclear. Given the TRS breaking nature of AMs, it is of particular interest to examine whether the thermal Hall conductivity is anomalous in these materials.

In this Article, we report systematic measurements of the THE in two representative altermagnet candidates, MnTe and CrSb. In semiconducting MnTe, where phonons dominate heat transport, we observe a pronounced anomalous thermal Hall response that far exceeds the electronic contribution estimated from the Wiedemann–Franz (WF) law. In metallic CrSb, after subtracting the electronic contribution, the remaining thermal Hall conductivity—primarily attributed to phonons—displays strong anomalous behavior. These results demonstrate the anomalous pTHE as a fundamental hallmark of altermagnetism, thus establishing phonon-mediated thermal transport as a superior probe to identify a new altermagnet—outperforming established electrical probing paradigms.

## Results

The $\alpha$-MnTe crystallizes in a NiAs-type structure ($P6_3$/mmc #194; Fig. 1a) and exhibits a compensated magnetic structure below the magnetic transition temperature $T_{AM} \approx 307$ K. The spins of $Mn^{2+}$ ions are aligned parallel within the *ab* plane and anti-parallel along the *c* axis. The two sublattices with opposite spin orientations are interconnected by a sixfold rotation or a mirror operation, defining the symmetry of this altermagnetic phase. Figure 1b shows the temperature-dependent magnetic susceptibility of a MnTe single crystal (photo in the inset) measured from 2.1 to 350 K under zero-field cooling (ZFC) condition, revealing clear magnetic anisotropy below $T_{AM}$, consistent with the in-plane Néel vector. Furthermore, the isothermal magnetization $M(H)$ curves for $B \parallel c$, presented in Fig. 1c, are approximately linear at various temperatures, exhibiting typical antiferromagnetic behavior without any canted component.

Before measuring the thermal Hall effect, we have characterized the electrical transport properties of MnTe single crystal, as shown in Figs. 2a,b. The resistivity $\rho_{xx}$ displays two distinct features: a sharp drop at $T_{AM}$, followed by an upturn below 60 K. The Hall conductivity $\sigma_{xy}$ is positive, indicating that MnTe behaves as a *p*-type semiconductor. However, $\sigma_{xy}$ varies linearly across the measured range, with no apparent anomalies, similar to some samples of Dai's group in ref. 25. Figure 2c presents the thermal conductivity $\kappa_{xx}$ which exhibits a pronounced phonon peak around 20 K. Notably, a thermal Hall signal has been observed in this material. The temperature-dependent total thermal Hall conductivity $\kappa_{xy}$ follows the phonon peak in $\kappa_{xx}$, suggesting that phonons dominate heat transport in MnTe, as in various other materials[37,41,43,45]. Unlike nonmagnetic and antiferromagnetic materials, $\kappa_{xy}$ in MnTe exhibits a peculiar magnetic field dependence (Fig. 2d). As the magnetic field increases, $\kappa_{xy}$ shows a hump around 1 T, followed by a sign reversal from positive to negative. This behavior is reminiscent of the electric anomalous Hall effect observed in many topological materials. The phonon thermal Hall conductivity $\kappa_{xy}^{ph}$ is extracted by subtracting the carrier contribution using the WF Law $\kappa/\sigma T = L_0$, where $L_0 = 2.45 \times 10^{-}$

[8] WΩK$^{-2}$ is the Lorentz ratio, as shown in Fig. 2e. Due to the relatively small contribution of hole carriers in MnTe (from –0.9% at 10 K to –17.3% at 60 K), $\kappa_{xy}^{ph}$ has a similar field dependence to the total thermal Hall conductivity. The phonon thermal Hall conductivity consists of two components: one is the conventional part, which is linear in magnetic field, akin to that previously observed in nonmagnetic or collinear antiferromagnetic materials, and the other is the anomalous part. We fit the data using the following function[48]:

$$\kappa_{xy}^{ph} = kB + \kappa_A \tanh(B/B_0). \quad (1)$$

The former linear term represents the conventional part, while the latter indicates the anomalous behavior, where $k$ is the slope of the linear term, $\kappa_A$ is the saturation value of the anomalous term, and $1/B_0$ is the differential permeability at $B$ = 0 T. After subtracting the linear term, the anomalous pTHE is extracted and presented in Fig. 2f, analogous to the THE in FMs[29,32,38].

To examine whether the anomalous pTHE is a general feature of AMs, we turn to another altermagnetic candidate CrSb, which has a magnetic structure similar to that of $\alpha$-MnTe and a transition temperature $T_{AM}$ above 700 K. However, the spins of $Cr^{2+}$ ions are aligned along the $c$ axis (see Fig. 3a). Figure 3b presents the temperature-dependent magnetic susceptibility of a CrSb single crystal, measured from 1.8 to 300 K under ZFC condition. These data reveal distinct magnetic anisotropy, consistent with the out-of-plane Néel vector. The isothermal magnetization curves $M(H)$ for $B \parallel c$, shown in Fig. 3c, exhibit a small hysteresis and a weak ferromagnetic component with a saturation field near $B$ = 0.1 T. This behavior differs from the out-of-plane response of MnTe but resembles its in-plane response, as shown in the Supplementary Fig. S2.

Similarly, we measured the electrical and thermal transport properties of the CrSb single crystal. As shown in Figs. 4a,b, $\rho_{xx}$ exhibits clear metallic behavior with a residual resistivity ratio ($RRR$) of 4.41, indicating high sample quality. The negative Hall conductivity $\sigma_{xy}$ is nonlinear in field across all temperatures, which was attributed to multiband effects[19–21], because the symmetry of CrSb forbids an anomalous Hall effect[7]. Figure 4c shows the longitudinal thermal conductivity $\kappa_{xx}$ of CrSb. After subtracting the electronic part using the WF law, the phonon thermal conductivity reveals a broad peak around 40 K confirming a substantial electronic contribution to heat transport in metallic CrSb, unlike the phonon-dominated transport in MnTe.

Figure 4d shows the total thermal Hall conductivity $\kappa_{xy}$ of CrSb, which mirrors the non-linear field dependence of the electrical Hall signal, suggesting an electronic origin at first glance. However, after subtracting the electronic thermal Hall contribution using WF law, there is still a clear phonon contribution with nonlinear field dependence. At $T = 80$ K and $B = 9$ T, the total, electronic and phonon thermal Hall conductivities are –194.17, –333.75 and 139.58 mW/(K m), respectively. The resulting phonon thermal Hall conductivity $\kappa_{xy}^{ph}$ is robust. Following the same procedure as for MnTe, subtracting a conventional linear-in-field background, we obtain the anomalous phonon thermal Hall conductivity plotted in Fig. 4f. It should be emphasized that this anomalous THE arises from a phonon contribution independent of charge carriers, closely resembling the behavior observed in MnTe. These findings provide compelling evidence for an intrinsic anomalous pTHE in CrSb, further supporting the presence of time-reversal symmetry breaking in AMs.

## Discussion

The pTHE has been widely observed in both magnetic and nonmagnetic materials. Except when the lattice thermal conductivity exhibits a strong magnetic-field dependence, the thermal Hall conductivity in the majority of materials is linear in magnetic field[34–37,39,41,44,45]. This conventional linear-in-field thermal Hall response is also evident in MnTe and CrSb. More importantly, after subtracting this conventional linear-in-field background, a sizable anomalous $\kappa_{xy}^{ph}$ remains. Its field dependence is reminiscent of anomalous thermal Hall responses previously observed in ferromagnets, such as $Lu_2V_2O_7$, $Fe_2Mo_3O_8$ and $VI_3$[29,32,38]. For ferromagnets, the anomalous thermal Hall responses typically track $M(H)$[29,32,38]. In altermagnets, very small ferromagnetic signals have been reported in $M(H)$ curves[49–51] (MnTe: in-plane; CrSb: out-of-plane), which are also present in our magnetization data (Supplementary Fig. S2 and Fig. 3c). However, the saturation field of these weak ferromagnetic components (~ 0.1 T) is an order of magnitude smaller than the field scale required to saturate the anomalous $\kappa_{xy}^{ph}$ ($\geq$ 1 T). This strong mismatch argues against weak net magnetization as the primary origin of the anomalous $\kappa_{xy}^{ph}$ in MnTe and CrSb.

Mechanisms tied to the Néel vector and crystalline symmetry therefore provide a more natural explanation. In altermagnets, direction-dependent superexchange originating from the specific arrangement of nonmagnetic ligands breaks the combined time-reversal/spatial-translation protections that otherwise suppress Berry curvature, yielding an intrinsic "crystal" thermal Hall effect carried by electrons or magnons[52,53], where the magnitude and sign of $\kappa_{xy}$ are determined by the Néel vector rather than by any small net magnetization. In our results, the electronic contribution has been subtracted according to WF law, and the anticipated magnon THE in ref. 53 is much smaller than the observed signal, therefore the only origin is from phonons.

The most plausible route is magnon–phonon hybridization. In collinear ordered magnets, such hybridization can generate chiral phonon modes[54–56], thereby transferring magnetic Berry curvature to the phonon sector. Even when phonons dominate heat transport, a large THE has been attributed to phonon–magnon hybridization[44,57]. Experimentally, magnon–phonon hybridization has been observed in altermagnetic $CoF_2$[58]. Besides the expected spin–phonon renormalization near the Néel temperature, a strong coupling between a one-magnon excitation and the lowest-frequency Raman-active $B_{1g}$ phonon was reported[58]. Although Raman scattering directly probes optical phonons, magnon–phonon coupling is not confined to zone-center modes. It can also occur for acoustic branches when energy and momentum conservation are satisfied, particularly near anticrossings between magnon and phonon dispersions[44,57]. Such hybrid quasiparticles carry finite phonon angular momentum suggesting phononic counterparts to the electronic response effects in altermagnets[59]. These findings provide the prerequisites for a hybridization-driven THE: phonon–magnon hybridization can imprint and substantially amplify an anomalous phonon $\kappa_{xy}$ in altermagnets. In this scenario, a Néel-vector-driven crystal THE does not require a large anomalous electrical Hall effect, as is the case with CrSb and MnTe[19–25].

Our observation of anomalous pTHE in altermagnets directly links the Néel vector to lattice vibrations, paving the way for acoustic manipulation of altermagnetic order. Because altermagnetic spin splitting is intrinsically tied to the crystal lattice[1,2], external strain[50,60]—whether static or dynamic—can efficiently control the Néel orientation and associated transport phenomena. Conversely, the observed magnon–phonon coupling implies that altering the magnetic state via acoustic waves will modify phonon spectra. This mutual control opens prospects for low-loss phononic devices, including acoustic-field-driven Néel switching and altermagnetic thermally memories that do not rely on charge currents.

In summary, we have measured THE of two altermagnet candidates, semiconducting $\alpha$-MnTe and metallic CrSb, and report the observation of a robust anomalous phonon $\kappa_{xy}$ in both materials after subtracting the carrier and conventional phonon contributions. The observed field scales and magnetization behavior rule out weak ferromagnetism as the primary cause and point toward a Néel-vector- and lattice-symmetry-driven origin, possibly mediated or amplified by magnon–phonon hybridization. Our results extend THE studies into altermagnets, highlight the important role phonons are playing, and motivate further theoretical and experimental work to identify the microscopic mechanism.

## Methods

**Samples growth.** MnTe single crystals were grown using an Sb-flux-assisted method. Starting materials, including Mn plates (99.9%, Macklin), Te powder (99.99%, Macklin), and Sb blocks (99.999%, Macklin), were placed in an alumina crucible with a molar ratio of Mn:Te:Sb = 1:1:20. To prevent oxidation, the crucible was sealed in a quartz ampoule under vacuum. The ampoule was then heated to 1050 °C over 15 h, held at that temperature for 10 h to ensure complete dissolution and homogenization, and subsequently cooled slowly at 2 °C/h to 700 °C. At this point, the excess Sb flux was efficiently removed by rapid centrifugation. The resulting black MnTe crystals exhibited shiny surfaces and typical dimensions of approximately 5 × 4 × 1 $mm^3$.

Single crystals of CrSb were grown using a chemical vapor transport method.

Stoichiometric amounts of chromium powder (Cr, 99.5%, Macklin) and antimony blocks (Sb, 99.999%, Macklin) were mixed in a 1:1 molar ratio, with 0.1 g of iodine granules ($I_2$, 99.99%, Macklin) added as the transport agent. The mixture was loaded into a 14-cm-long quartz ampoule, which was then evacuated and sealed under vacuum. The sealed ampoule was placed horizontally in a two-zone tube furnace, with the source zone heated to 850 °C and the growth zone to 750 °C over 10 h. This temperature gradient was maintained for 150 h to facilitate crystal growth, after which the furnace was turned off and allowed to cool to room temperature. The resulting CrSb crystals were carefully extracted from the ampoule and rinsed with ethanol to remove residual iodine from their surfaces. Lustrous black plate-like crystals with typical dimensions of approximately 9 × 7 × 1 $mm^3$ were obtained.

**Measurements.** The VSM magnetization measurement was performed down to 2.1 K using a magnetic property measurement system (MPMS, Quantum Design). Samples for resistivity and thermal conductivity measurements were cut and polished into nearly rectangular shapes. The dimensions were 3.57 × 1.43 × 0.15 $mm^3$ for MnTe and 1.52 × 1.14 × 0.11 $mm^3$ for CrSb, respectively. The electrical and thermal transport measurements were conducted in a Physical Property Measurement system (PPMS, Quantum Design) under high-vacuum environment. The heater and thermometers were connected to the sample by silver wires. The contacts were made of silver paint for MnTe and silver epoxy annealed at 120 °C for CrSb, respectively. Three-thermometers (Cernox 1050) method was employed to simultaneously measure the longitudinal and transverse thermal gradients. A constant heat current $Q$ was applied along the $x$ axis lying in the basal plane of the single crystal, using a resistive heater connected to one end of the sample, generating a longitudinal temperature difference $\Delta T_x = T_1 - T_2$ (Supplementary Fig. S1). The other end of the sample was attached to a heat sink with silver paint. The copper block is used as heat sink. The thermal conductivity along the $x$ axis was calculated as $\kappa_{xx} = (Q/\Delta T_x)(L/wt)$. By applying a magnetic field $B$ along the $z$ directions, a transverse gradient $\Delta T_y = T_3 - T_2$ was generated. The thermal Hall conductivity was defined as $\kappa_{xy} = \kappa_{yy}(\Delta T_y/\Delta T_x)(L/w)$, where $\kappa_{yy}$ is the longitudinal thermal conductivity along the $y$ axis. We assumed $\kappa_{yy} = \kappa_{xx}$ for crystals with high symmetry. The contamination from $\kappa_{xx}$ in $\kappa_{xy}$ due to contact misalignment is removed by performing field anti-symmetrization of the transverse temperature difference $\Delta T_y$, i.e. $\Delta T_y = [\Delta T_y(H) - \Delta T_y(-H)]/2$. The positive direction of magnetic field is illustrated by blue arrows in Supplementary Fig. S1.

## Data availability

The data that support the findings of this study are available from the corresponding author upon reasonable request.

## References

1. Šmejkal, L., Sinova, J. & Jungwirth, T. Beyond Conventional Ferromagnetism and Antiferromagnetism: A Phase with Nonrelativistic Spin and Crystal Rotation Symmetry. *Phys. Rev. X* **12**, 031042 (2022).
2. Šmejkal, L., Sinova, J. & Jungwirth, T. Emerging Research Landscape of Altermagnetism. *Phys. Rev. X* **12**, 040501 (2022).
3. Song, C. *et al.* Altermagnets as a new class of functional materials. *Nat. Rev. Mater.* **10**, 473 (2025).
4. Jungwirth, T. *et al.* Symmetry, microscopy and spectroscopy signatures of altermagnetism. *Nature* **649**, 837 (2026).
5. Ma, H.-Y. *et al.* Multifunctional antiferromagnetic materials with giant piezomagnetism and noncollinear spin current. *Nat. Commun.* **12**, 2846 (2021).
6. Liu, P., Li, J., Han, J., Wan, X. & Liu, Q. Spin-Group Symmetry in Magnetic Materials with Negligible Spin-Orbit Coupling. *Phys. Rev. X* **12**, 021016 (2022).
7. Šmejkal, L., González-Hernández, R., Jungwirth, T. & Sinova, J. Crystal time-reversal symmetry breaking and spontaneous Hall effect in collinear antiferromagnets. *Sci. Adv.* **6**, eaaz8809 (2020).
8. Feng, Z. *et al*. An anomalous Hall effect in altermagnetic ruthenium dioxide. *Nat. Electron.* **5**, 735 (2022).
9. Liu, J. *et al*. Absence of Altermagnetic Spin Splitting Character in Rutile Oxide $RuO_2$. *Phys. Rev. Lett.* **133**, 176401 (2024).
10. Ding, J. *et al*. Large Band Splitting in *g*-Wave Altermagnet CrSb. *Phys. Rev. Lett.* **133**, 206401 (2024).
11. Reimers, S. *et al*. Direct observation of altermagnetic band splitting in CrSb thin films. *Nat. Commun.* **15**, 2116 (2024).
12. Li, C. *et al*. Topological Weyl altermagnetism in CrSb. *Commun. Phys.* **8**, 311 (2024).
13. Liao, S. *et al*., Direct Observation of Large Altermagnetic Splitting in CrSb (100) Thin Film. *Chin. Phys. Lett.* **42**, 067503 (2025).
14. Krempaský, J. *et al*. Altermagnetic lifting of Kramers spin degeneracy. *Nature* **626**, 517 (2024).
15. Lee, S. *et al*. Broken Kramers Degeneracy in Altermagnetic MnTe. *Phys. Rev. Lett.* **132**, 036702 (2024).
16. Liu, Z., Ozeki, M., Asai, S., Itoh, S. & Masuda, T. Chiral Split Magnon in Altermagnetic MnTe. *Phys. Rev. Lett.* **133**, 156702 (2024).
17. Osumi, T. *et al.* Observation of a giant band splitting in altermagnetic MnTe. *Phys. Rev. B* **109**, 115102 (2024).
18. Zhang, T. *et al*. Evidence for Itinerant Electron-Local Moment Interaction in Li-Doped $\alpha$-MnTe. *arXiv*: 2512.00747 (2025).
19. Urata, T., Hattori, W. & Ikuta, H. High mobility charge transport in a multicarrier altermagnet CrSb. *Phys. Rev. Mater.* **8**, 084412 (2024).
20. Bai, Y. *et al*. Nonlinear field dependence of Hall effect and high-mobility multi-carrier transport in an altermagnet CrSb. *Appl. Phys. Lett.* **126**, 042402 (2025).

21. Peng, X. *et al*. Scaling behavior of magnetoresistance and Hall resistivity in the altermagnet CrSb. *Phys. Rev. B* **111**, 144402 (2025).
22. Wasscher, J. D. Evidence of weak ferromagnetism in MnTe from galvanomagnetic measurements. *Solid State Commun.* **3**, 169 (1965).
23. Gonzalez Betancourt, R. D. *et al*. Spontaneous Anomalous Hall Effect Arising from an Unconventional Compensated Magnetic Phase in a Semiconductor. *Phys. Rev. Lett.* **130**, 036702 (2023).
24. Kluczyk, K. P. *et al*. Coexistence of anomalous Hall effect and weak magnetization in a nominally collinear antiferromagnet MnTe. *Phys. Rev. B* **110**, 155201 (2024).
25. Liu, Z. *et al*. Strain-Tunable Anomalous Hall Effect in Hexagonal MnTe. *arXiv*: 2509.19582 (2025).
26. Tschirner, T. *et al*. Saturation of the anomalous Hall effect at high magnetic fields in altermagnetic $RuO_2$. *APL Mater.* **11**, 101103 (2023).
27. Jeong, S. G. *et al*. Metallicity and anomalous Hall effect in epitaxially strained, atomically thin $RuO_2$ films. *Proc. Natl. Acad. Sci. U.S.A.* **122**, e2500831122 (2025).
28. Strohm, C., Rikken, G. L. J. A. & Wyder, P. Phenomenological Evidence for the Phonon Hall Effect. *Phys. Rev. Lett.* **95**, 155901 (2005).
29. Onose, Y. *et al*. Observation of the Magnon Hall Effect. *Science* **329**, 297 (2010).
30. Hirschberger, M., Chisnell, R., Lee, Y. S. & Ong, N. P. Thermal Hall Effect of Spin Excitations in a Kagome Magnet. *Phys. Rev. Lett.* **115**, 106603 (2015).
31. Hirschberger, M., Krizan, J. W., Cava, R. J. & Ong, N. P. Large thermal Hall conductivity of neutral spin excitations in a frustrated quantum magnet. *Science* **348**, 106 (2015).
32. Ideue, T., Kurumaji, T., Ishiwata, S. & Tokura, Y. Giant thermal Hall effect in multiferroics. *Nat. Mater.* **16**, 797 (2017).
33. Kasahara, Y. *et al*. Majorana quantization and half-integer thermal quantum Hall effect in a Kitaev spin liquid. *Nature* **559**, 227 (2018).
34. Grissonnanche, G. *et al*. Giant thermal Hall conductivity in the pseudogap phase of cuprate superconductors. *Nature* **571**, 376 (2019).
35. Grissonnanche, G. *et al*. Chiral phonons in the pseudogap phase of cuprates. *Nat. Phys.* **16**, 1108 (2020).
36. Boulanger, M.-E. *et al*. Thermal Hall conductivity in the cuprate Mott insulators $Nd_2CuO_4$ and $Sr_2CuO_2Cl_2$. *Nat. Commun.* **11**, 5325 (2020).
37. Li, X., Fauqué, B., Zhu, Z. & Behnia, K. Phonon Thermal Hall Effect in Strontium Titanate. *Phys. Rev. Lett.* **124**, 105901 (2020).
38. Zhang, H. *et al*. Anomalous Thermal Hall Effect in an Insulating van der Waals Magnet. *Phys. Rev. Lett.* **127**, 247202 (2021).
39. Chen, L., Boulanger, M.-E., Wang, Z.-C., Tafti, F. & Taillefer, L. Large phonon thermal Hall conductivity in the antiferromagnetic insulator $Cu_3TeO_6$. *Proc. Natl. Acad. Sci. U.S.A.* **119**, e2208016119 (2022).
40. Gillig, M. *et al*. Phononic-magnetic dichotomy of the thermal Hall effect in the Kitaev material $Na_2Co_2TeO_6$. *Phys. Rev. Res.* **5**, 043110 (2023).
41. Li, X. *et al*. The phonon thermal Hall angle in black phosphorus. *Nat. Commun.* **14**, 1027 (2023).

42. Ataei, A. *et al*. Phonon chirality from impurity scattering in the antiferromagnetic phase of $Sr_2IrO_4$. *Nat. Phys.* **20**, 585–588 (2024).
43. Sharma, R., Valldor, M. & Lorenz, T. Phonon thermal Hall effect in nonmagnetic $Y_2Ti_2O_7$. *Phys. Rev. B* **110**, L100301 (2024).
44. Meng, Q. *et al*. Thermodynamic Origin of the Phonon Hall Effect in a Honeycomb Antiferromagnet. *arXiv*:2403.13306 (2024).
45. Jin, X. B. *et al*. Discovery of Universal Phonon Thermal Hall Effect in Crystals. *Phys. Rev. Lett.* **135**, 196302 (2025).
46. Behnia, K. Phonon thermal Hall as a lattice Aharonov-Bohm effect. *SciPost Phys. Core* **8**, 061 (2025).
47. Oh, T. Thermal Hall Effect Induced by Phonon Skew-Scattering via Orbital Magnetization. *arXiv*:2507.22436 (2025).
48. Chen, X., Xie, H., Shen, H. & Wu, Y. Vector Magnetometer Based on a Single Spin-Orbit-Torque Anomalous-Hall Device. *Phys. Rev. Appl.* **18**, 024010 (2022).
49. Kim, W. *et al*. Room-Temperature Ferromagnetic Property in MnTe Semiconductor Thin Film Grown by Molecular Beam Epitaxy. *IEEE Trans. Magn.* **45**, 2424 (2009).
50. Zhou, Z. *et al*. Manipulation of the altermagnetic order in CrSb via crystal symmetry. *Nature* **638**, 645 (2025).
51. Tseng, C.-H. *et al*. Epitaxial Growth of Altermagnet CrSb via Magnetron Sputtering. *Cryst. Growth Des.* **25**, 9171 (2025).
52. Zhou, X. *et al*. Crystal Thermal Transport in Altermagnetic $RuO_2$, *Phys. Rev. Lett.* **132**, 056701 (2024).
53. Hoyer, R., Jaeschke-Ubiergo, R., Ahn, K.-H., Šmejkal, L. & Mook, A. Spontaneous crystal thermal Hall effect in insulating altermagnets. *Phys. Rev. B* **111**, L020412 (2025).
54. Nomura, T. *et al.* Phonon Magnetochiral Effect. *Phys. Rev. Lett.* **122**, 145901 (2019).
55. Thingstad, E., Kamra, A., Brataas, A. & Sudbø, A. Chiral Phonon Transport Induced by Topological Magnons. *Phys. Rev. Lett.* **122**, 107201 (2019).
56. Cui, J. *et al.* Chirality selective magnon-phonon hybridization and magnon-induced chiral phonons in a layered zigzag antiferromagnet. *Nat. Commun.* **14**, 3396 (2023).
57. Yang, H., Go, G., Park, J., Kim, S. K. & Park, J.-G. Exchange striction induced thermal Hall effect in the van der Waals antiferromagnet $MnPS_3$. *Phys. Rev. B* **110**, 165147 (2024).
58. Prosnikov, M. A., Bal, M., Pisarev, R. V., Christianen, P. C. M. & Kalashnikova, A. M. Giant Intrinsic Nonlinear Phonon-Magnon Coupling in the Antiferromagnet $CoF_2$. *arXiv*:2405.07304 (2024).
59. Bendin, H., Mook, A., Mertig, I. & Neumann, R. R. D-Wave Phonon Angular Momentum Texture in Altermagnets by Magnon-Phonon-Hybridization. *arXiv*:2511.08357 (2025).
60. Aoyama, T. & Ohgushi, K. Piezomagnetic properties in altermagnetic MnTe. *Phys. Rev. Mater.* **8**, L041402 (2024).

## Acknowledgements

This work is supported by the Natural Science Foundation of China (Grant No. 12534004), the Shanghai Municipal Science and Technology Major Project (Grant No. 2019SHZDZX01), and Quantum Science and Technology-National Science and Technology Major Project (Grant No. 2024ZD0300104). Y.F.G. acknowledges the National Key R&D Program of China (Grants No. 2024YFA1408400 and 2023YFA140610) and the open research fund of Beijing National Laboratory for Condensed Matter Physics (2023BNLCMPKF002).

## Author Contributions

S.Y.L. and X.Z. conceived the idea and designed the experiments. Y.X.L and Y.F.G. grew the MnTe and CrSb samples. W.B.W. and X.Z. performed the measurements. W.B.W., X.Z. and S.Y.L. analyzed the data. W.B.W, X.Z. and S.Y.L. wrote the paper with assistance from all the authors. W.B.W, X.Z. contributed equally to this work.

## Competing interests

The authors declare no competing interests.

## Additional Information

**Supplementary information** is available for this paper at URL inserted when published.

**Correspondence** and requests for materials should be addressed to S.Y.L. (shiyan_li@fudan.edu.cn) and X.Z. (xuzhang_fd@fudan.edu.cn).

Figure 1

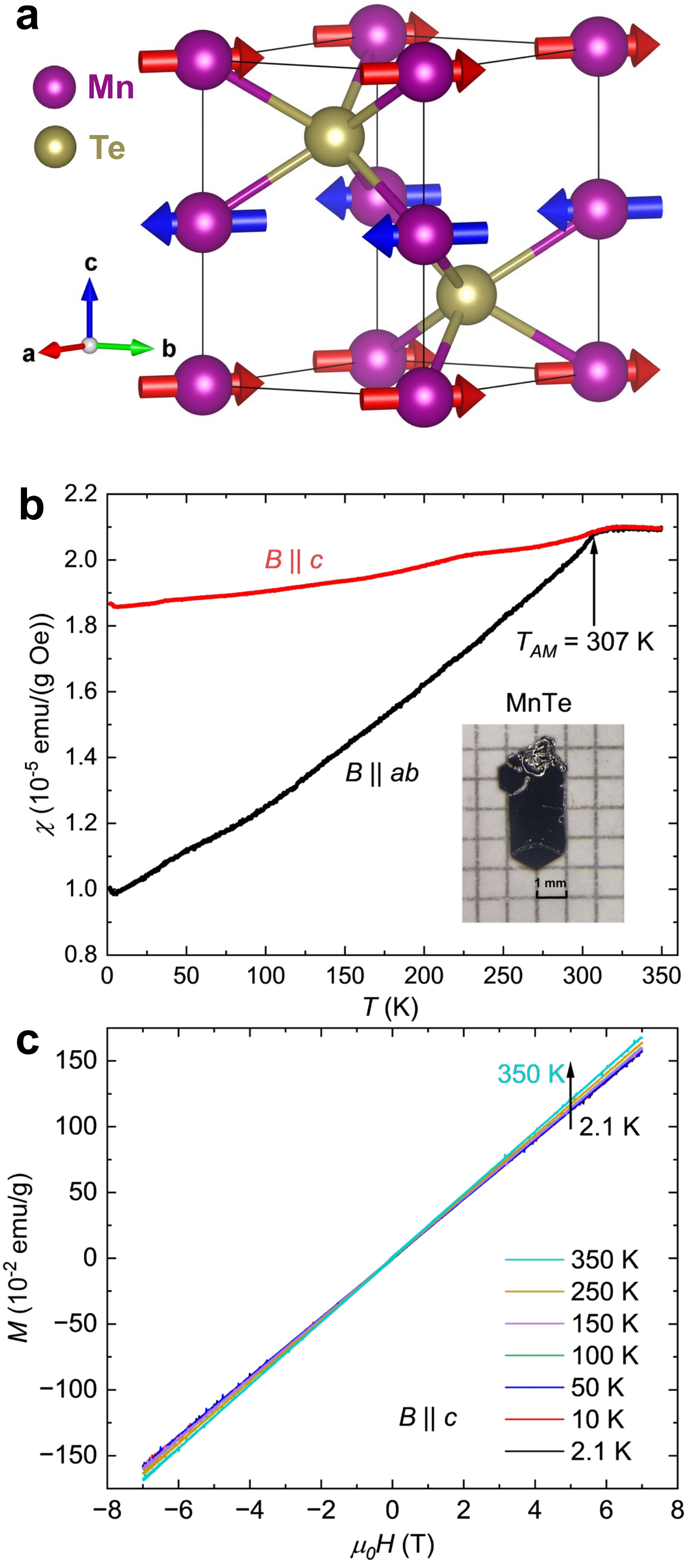

# Figure 2

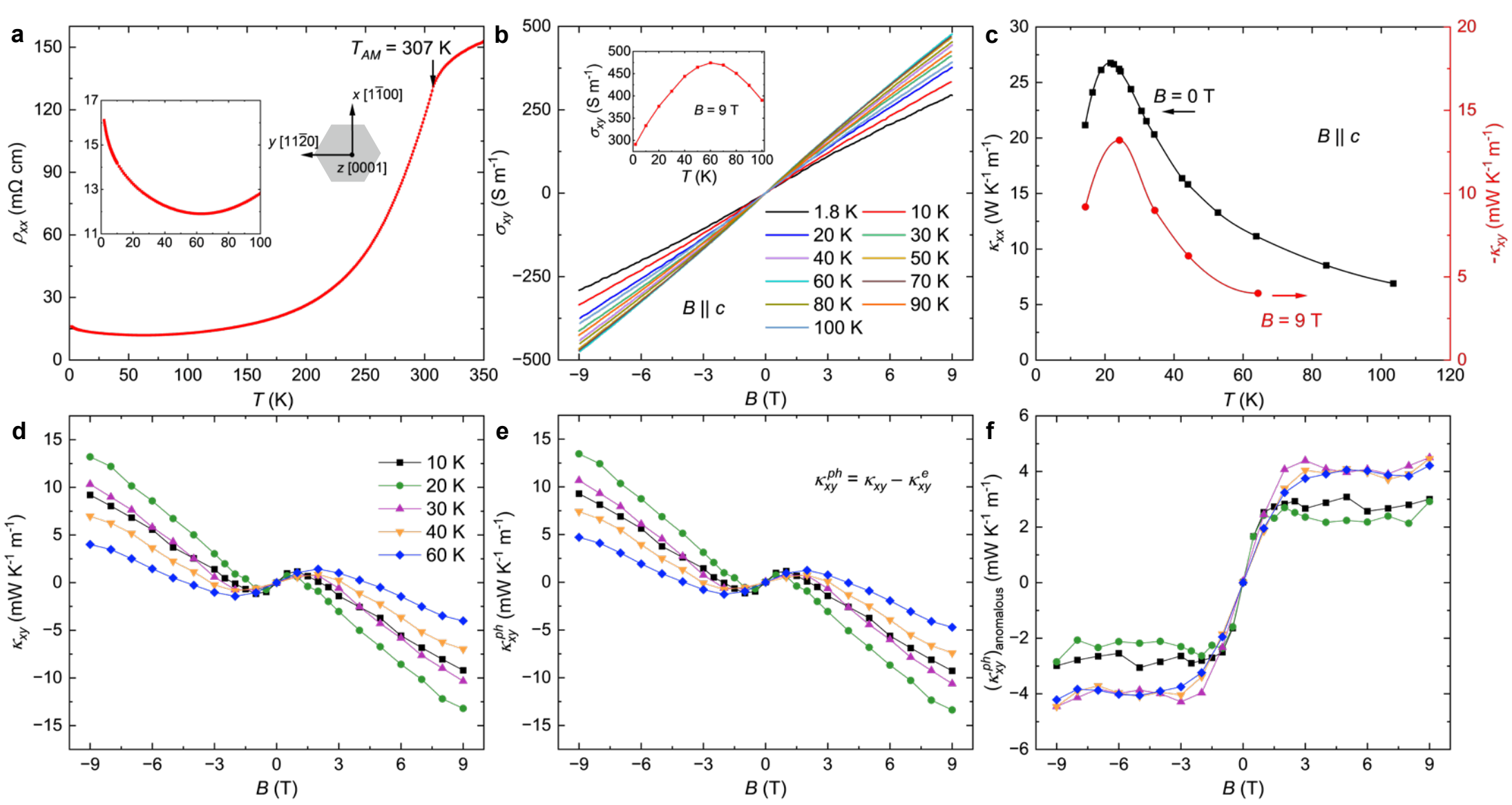

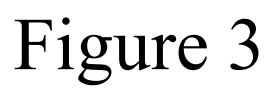

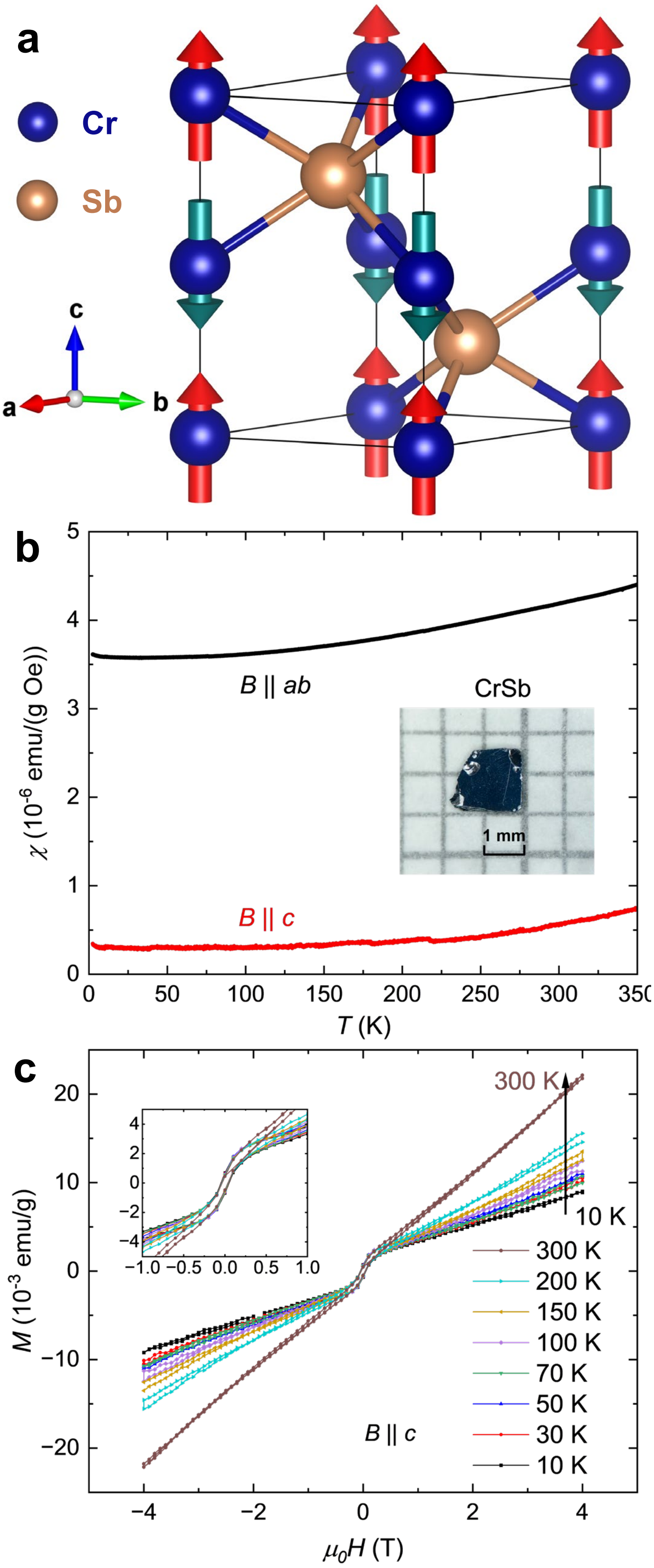

# Figure 4

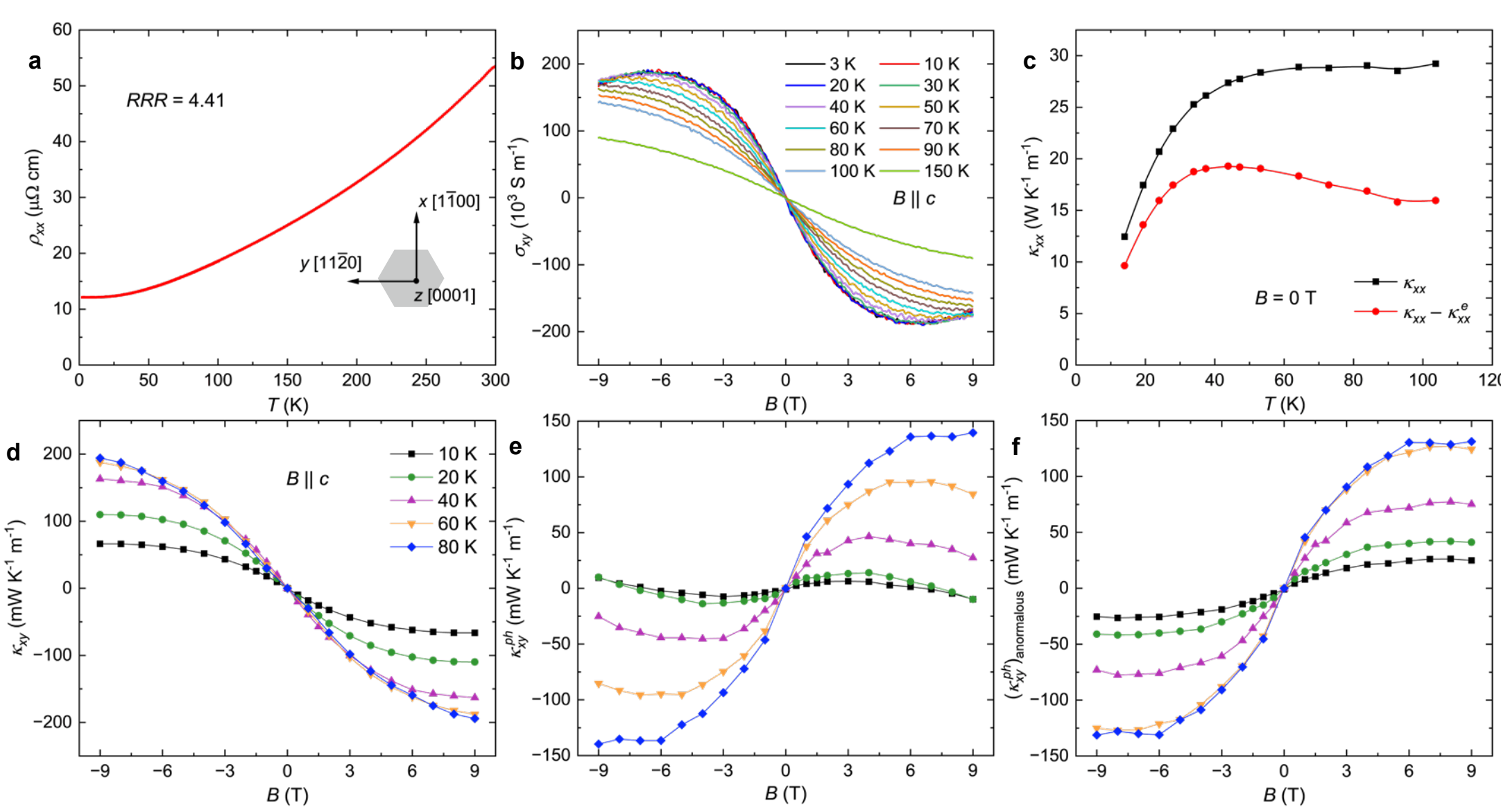

## Figure captions

**Figure 1 | Crystal structure and magnetization of *α*-MnTe. a**, The crystal structure of *α*-MnTe with opposite-spin sublattices. **b**, Temperature-dependent magnetic susceptibility measured under ZFC conditions for different magnetic-field directions. The inset shows an optical image of the MnTe crystal used for the measurements. **c**, Field-dependent magnetization at different temperatures for $B \parallel c$.

**Figure 2 | Electrical and thermal transport in *α*-MnTe. a**, In-plane longitudinal resistivity of a *α*-MnTe single crystal in zero field. The left inset shows an enlarged view of the low-temperature region. The right inset defines specific crystalline orientations in the hexagonal crystal. **b**, Field-dependent Hall conductivity at different temperatures for $B \parallel c$. The inset presents the temperature-dependence of Hall conductivity at $B = 9$ T. **c**, Temperature-dependent longitudinal ($B = 0$ T) and transverse ($B = 9$ T) thermal conductivities. **d, e, f** Field-dependent total, phonon and anomalous thermal Hall conductivities at different temperatures for $B \parallel c$.

**Figure 3 | Crystal structure and magnetization of CrSb. a**, The crystal structure of hexagonal CrSb with opposite-spin sublattices. **b**, Temperature-dependent magnetic susceptibility under ZFC conditions for different magnetic-field directions. The inset shows an optical image of the CrSb crystal used for experiments. **c**, Field-dependent magnetization at various temperatures for $B \parallel c$. The inset presents an enlarged view of the low-field region.

**Figure 4 | Electrical and thermal transport in CrSb. a**, In-plane longitudinal resistivity of a CrSb single crystal in zero field. The inset defines specific crystalline orientations in the hexagonal crystal. (b) Field-dependent Hall conductivity at different temperatures for $B \parallel c$. (c) Temperature-dependent total and phonon-only longitudinal thermal conductivities at $B = 0$ T. **d, e, f** Field-dependent total, phonon and anomalous thermal Hall conductivities at different temperatures for $B \parallel c$.

Supplementary Information for

# “Observation of anomalous thermal Hall effect in altermagnets”

Wenbo Wan [1,2*], Xu Zhang[1*✉], Yixuan Luo[3], Yanfeng Guo[3,4✉] and Shiyan Li[1,2,5,6✉]

[1]*State Key Laboratory of Surface Physics, and Department of Physics, Fudan University, Shanghai 200438, China*

[2]*Shanghai Research Center for Quantum Sciences, Shanghai 201315, China*

[3]*State Key Laboratory of Quantum Functional Materials, School of Physical Science and Technology, ShanghaiTech University, Shanghai 201210, China*

[4]*ShanghaiTech Laboratory for Topological Physics, ShanghaiTech University, Shanghai 201210, China*

[5]*Shanghai Branch, Hefei National Laboratory, Shanghai 201315, China*

[6]*Collaborative Innovation Center of Advanced Microstructures, Nanjing 210093, China*

Corresponding author. Email: shiyan_li@fudan.edu.cn (S.Y.L.); guoyf@shanghaitech.edu.cn (Y.F.G.); xuzhang_fd@fudan.edu.cn (X.Z.)

## Supplementary Note 1: Setup of thermal transport measurement.

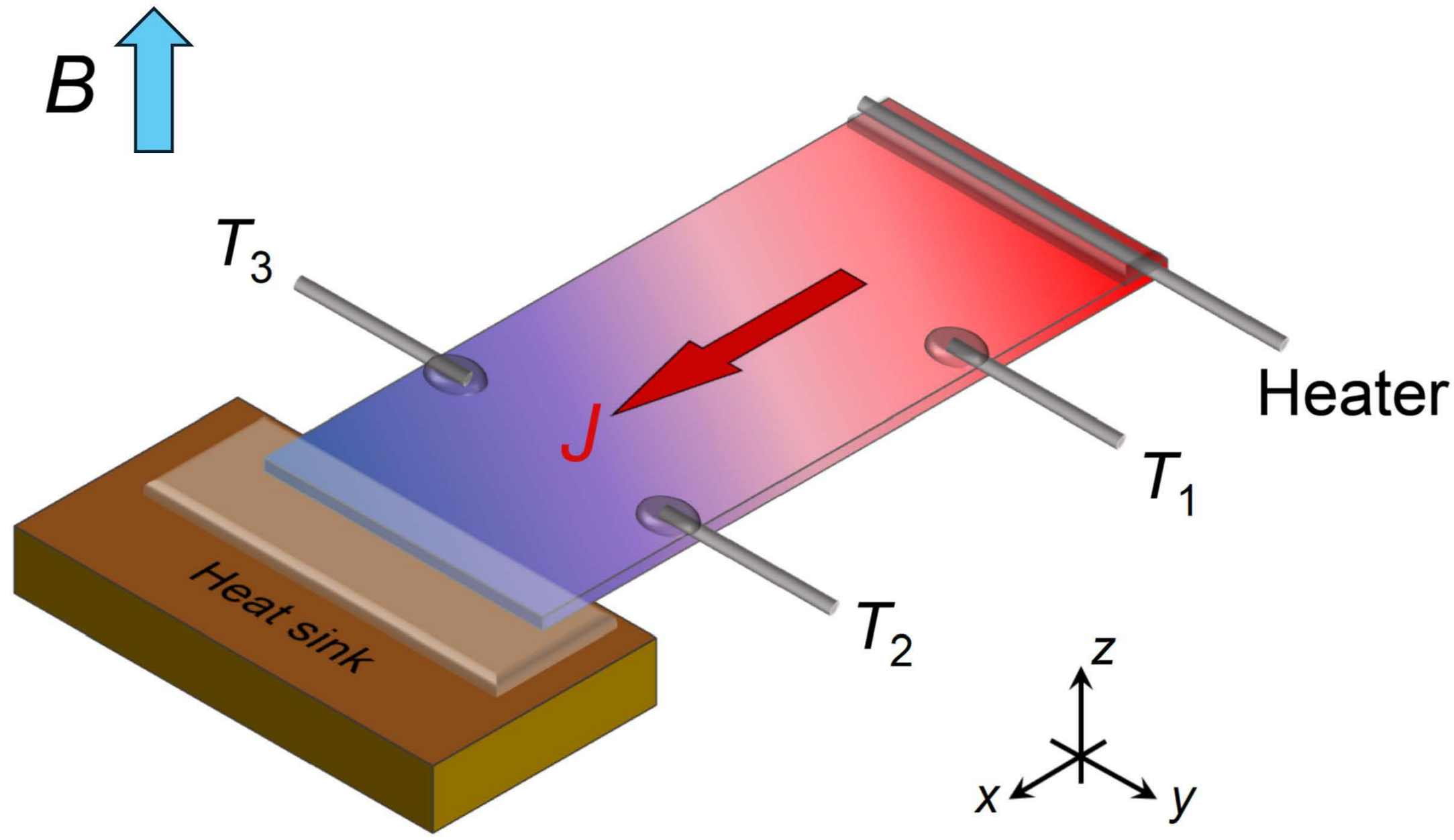

**Figure S1 | Setup of thermal transport measurement.**

## Supplementary Note 2: In-plane *M*(*H*) curves of MnTe.

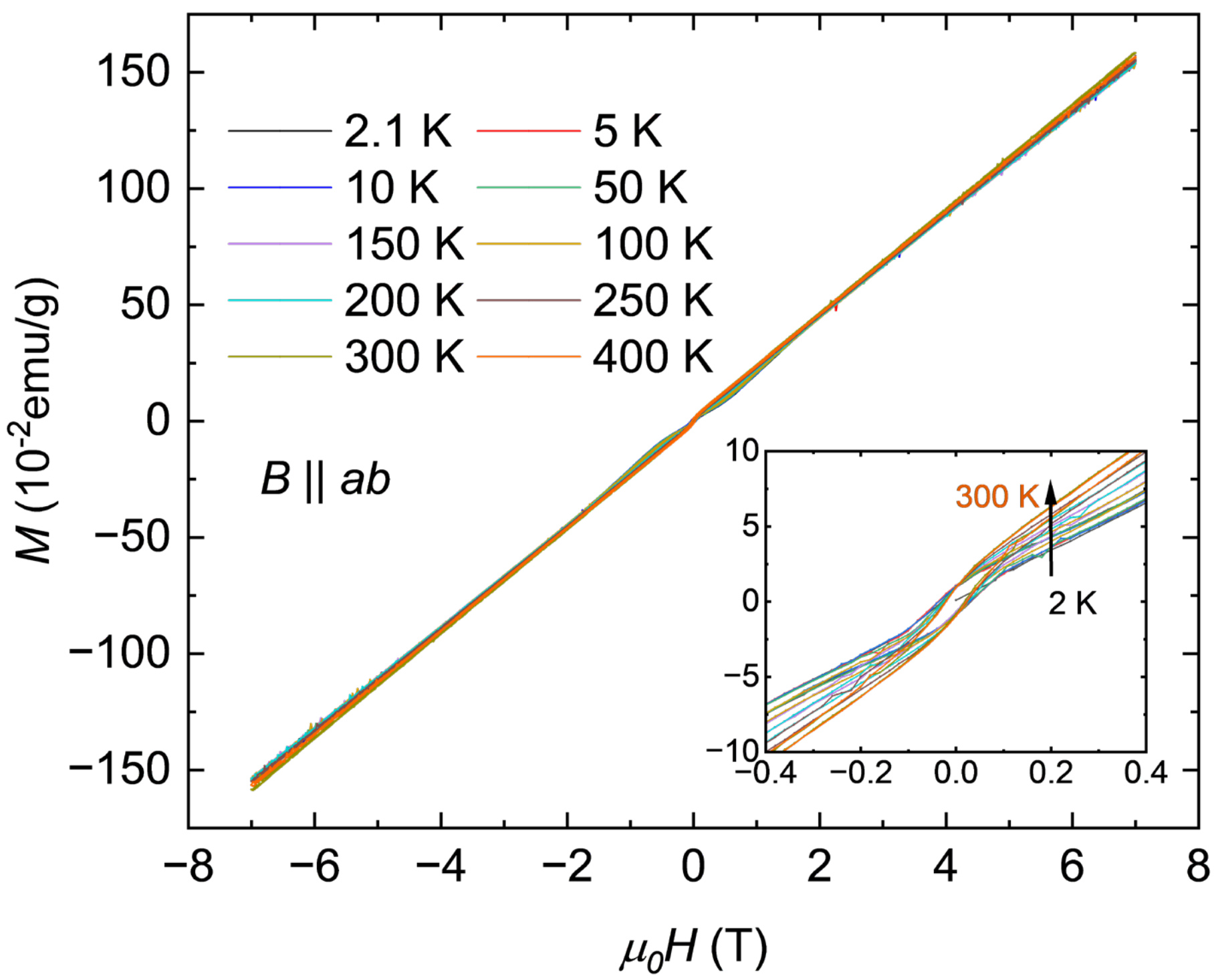

**Figure S2 | In-plane *M*(*H*) curves of MnTe.** Field-dependent magnetization of MnTe at different temperatures for $B \parallel ab$. The inset presents an enlarged view of the low-field region.